\documentclass[12pt,preprint]{aastex}

\slugcomment{To appear in ApJ}

\shorttitle{Exploring brown dwarf disks}
\shortauthors{Scholz et al.}

\begin{document}


\title{Exploring brown dwarf disks: A 1.3\,mm survey in Taurus}


\author{Alexander Scholz, Ray Jayawardhana}
\affil{Department of Astronomy \& Astrophysics, University of Toronto,
    60 St. George Street, Toronto, Ontario M5S3H8, Canada}
\email{aleks@astro.utoronto.ca}
\author{Kenneth Wood}
\affil{School of Physics \& Astronomy, University of St. Andrews, North Haugh, 
St. Andrews KY16 9SS}
\email{kw25@st-andrews.ac.uk}

\begin{abstract}
We have carried out sensitive 1.3\,mm observations of 20 young brown dwarfs in
the Taurus star-forming region, representing the largest sample of
young substellar objects targeted in a deep millimeter continuum survey to date.
Under standard assumptions, the masses of brown dwarf disks range from
$\lesssim$0.4 to several Jupiter masses. Their {\em relative} disk masses are
comparable to those derived for coeval low-mass stars: most of them are in
the $\lesssim$1\% -- 5\% range, and there is no clear change of relative disk mass
with object mass from 0.015 to 3 solar masses. Specifically, we do not find evidence 
for disk truncation, as would be expected in the ejection scenario for brown dwarf 
origin, although the signature of ejection may be hidden in our non-detections.
We use the derived mm fluxes, complemented by mid-infrared data from the Spitzer Space
Telescope and ground-based near-infrared images, to construct spectral
energy distributions (SEDs) for six of our sources, and model those SEDs with 
a Monte Carlo radiative transfer code. While the model fits are by no means
unique, they allow us to investigate disk properties such as the degree of
flaring and minimum radii. In several cases, we find that the SEDs in the 
mid-infrared exhibit lower flux levels than predicted by hydrostatic models, 
implying dust settling to the disk midplane. What's more, at least 25\% of 
our targets are likely to have disks with radii $>$10 AU; models with smaller 
disks cannot reproduce the mm fluxes even if they are very massive. This 
finding is in contrast to the results of some simulations of the ejection 
scenario for brown dwarf formation that suggest only $\sim$5\% of ejected 
objects would harbor disks larger than 10\,AU. Our findings imply that 
ejection is probably not the dominant formation process, but may still be 
relevant for some brown dwarfs.
\end{abstract} 





\section{Introduction}
\label{intro}

More than a decade after the discovery of the first bona fide brown dwarfs
\citep{nok95,rzm95}, it is now firmly established that 
these objects with substellar masses ($M<0.08\,M_{\odot}$) are ubiquitous in star 
forming regions, open clusters, and the field \citep[see e.g.][]{cnk00,mbs03,krl00}. 
Several hundred brown dwarfs have been identified, clearly demonstrating 
that substellar objects bridge the mass range between stars and planets -- 
hence, the mass function is continuous from solar down to Jupiter-like masses.

This challenges the conventional understanding of the formation of stars and
planets: Whereas stars form through fragmentation and collapse of molecular
cloud cores, planets are believed to originate in subsequent processes in the
circumstellar disk. This implies that the formation process is a function of
object mass, and has led to a debate about the origin of brown dwarfs. Four
main scenarios have been discussed recently as possible sources of (isolated) 
brown dwarfs \citep[see][]{wg05}: a) Collapse of molecular cloud 
cores with substellar masses, i.e. {\it in situ} formation, a process comparable to 
the formation of stars \citep{pn04}. b) Planet-like formation in a circumstellar disk, 
followed by ejection \citep{pdc00}. c) Formation as stellar embryos in multiple 
systems that are ejected in an early stage \citep{rc01}. d) Photoevaporation of 
intermediate-mass cores \citep{wz04}. It became clear, however, that only star-like 
formation (a) and ejection from multiple systems (c) are able to produce significant 
numbers of brown dwarfs, and are thus considered to be the main scenarios for brown 
dwarf formation \citep{wg05,kb03}. 

Distinguishing between the two models has been a main motivation for observational
studies of young brown dwarfs. Soon it was apparent that substellar objects with
ages of a few Myrs share many properties with solar-mass T Tauri stars. Particularly,
near- and mid-infrared surveys clearly indicate the existence of 
circum-sub-stellar material around young brown dwarfs \citep[e.g.][]{mal01,jas03}, 
where the SEDs are well-described by models of either flat or flared 
accretion disks \citep{ntc02,pah03,mjn04,akc06}.
Additional evidence for the existence of substellar disks comes
from spectroscopic accretion studies: A significant fraction of young brown dwarfs
shows spectroscopic signatures of ongoing accretion and mass outflow, typical for 
classical T Tauri stars \citep{fc01,jmb03,mhc03,mjb05}. The main conclusion so far 
is that accretion disks around brown dwarfs are comparable to or at least not vastly 
different from stellar disks, in terms of their geometry, their accretion behaviour, 
and their lifetime \citep[e.g.][]{jmb02,bm03,sj06}. 

This finding alone is not sufficient to distinguish between the competing 
formation models: The pure existence of circum-sub-stellar disks does not rule
out an ejection, because simulations show that a substantial fraction of material
can survive the ejection process \citep{bbb02}. Unfortunately, quantitative testable 
predictions for the amount of dust and gas remaining after a typical encounter in a 
multiple system, which leads to the ejection of the lower mass body, are rare in the 
literature. \citet{h95} estimates the average mass loss through an encounter to be less 
than $\sim 50$\% of the initial disk mass, where most of the lost material may be captured
by the perturber. Is has also been predicted that brown dwarfs with disk radii 
larger than 10-20\,AU are rare \citep[$\sim 5$\%,][]{bbb02,bbb03}. Generally, it is 
believed that in a statistical sense an ejection process will significantly reduce
the disk mass and the disk radius of the ejected body \citep{rc01}, and thus leads 
to truncated disks. This provides motivation for studies of disk properties for brown 
dwarfs.

A further reason to explore disk masses in the substellar regime is the unsettled
issue of a possible trend of disk mass, absolute or relative, with object mass. Intuitively,
one expects lower-mass stars to have lower absolute disk masses, resulting
in a constant disk mass to object mass ratio, and indeed this has been found by some
authors \citep[see the review by][]{ngm00}. Other groups claim to find constant absolute
disk masses and, as a consequence, higher relative disk masses for lower mass stars
\citep[e.g.][]{ncz97,ms00}. The main problem, which might
prevent the detection of a clear trend, is the large scatter of 2-3 orders of magnitude 
in the measured disk masses at any given stellar mass. Extending the mass range to 
substellar objects might help to determine whether there is a correlation of disk with 
object mass or not.

The best way to determine disk masses and outer radii is to analyse SEDs with coverage from
near-infrared to the submillimeter or millimeter regime. Infrared SEDs alone are not 
sensitive to constrain these parameters \citep[see][]{akc06}. Moreover, the submm/mm flux 
is directly related to the dust mass in the disk, providing a straightforward way to estimate
disk masses. This method has been established by \citet{bsc90} for young stellar objects,
and comprehensive submm/mm surveys have been carried out for large samples
of T Tauri and Herbig Ae/Be stars \citep[e.g.][]{ob95,ser00,aw05}. The same method 
has been applied to a small sample of brown dwarfs by \citet{kap03}. Their initial 
study provided disk masses for two and upper limits for seven young brown dwarfs. 

Submm/mm observations of brown dwarfs are at the limit of the observational capabilities 
of current submm/mm telescopes, with typical flux levels lower than 5\,mJy, and require 
substantial observational efforts. Here we describe the first comprehensive study of disks
around very young substellar objects, aimed to probe brown dwarf formation theories
and disk properties as a function of object mass, and thus providing an observational
foundation for future theoretical studies of these problems. Our targets are 20 brown dwarfs
in the Taurus star forming region, which represent one of the largest coeval samples of brown
dwarfs known to date. For all 20 objects, we obtained deep integrations with the 1.3\,mm
bolometer camera at the IRAM 30\,m telescope. Six sources were detected at flux levels
between 2 and 8\,mJy at 1.3\,mm., thus increasing the number of brown dwarfs with known
disk masses by a factor of 4. To probe the disk geometry, we additionally made use of 
Spitzer mid-infrared data for objects detected at 1.3\,mm. By combining mm and mid-infrared
photometry and comparing with SED models, we aimed to constrain disk properties. Particularly,
we are interested in the disk radii, which can, similar to disk masses, be used to search for 
signatures of truncated disks and thus distinguish between formation models.

The paper is structured as follows. Sect. \ref{obs} contains a description of target sample,
observations, data reduction, and reliability checks. In Sect. \ref{diskmass} we discuss the
conversion from mm fluxes to disk masses, and compare our results with disk masses for stars
from the literature. The SED modeling based on mm and Spitzer data is described
in Sect. \ref{models}. The final Sect. \ref{sum} provides a summary of our results.

\section{Targets, observations, fluxes}
\label{obs}

This paper is based on a 1.3\,mm continuum survey of 20 brown dwarfs in the Taurus star forming
region. At the time of the observations, these objects were the only spectroscopically
confirmed Taurus members with spectral types later than M6 and thus most likely substellar
masses (see Sect. \ref{dm2} for a more detailed assessment of the object masses). The
targets were identified in photometric surveys with follow-up optical spectroscopy 
by \citet{mdm01}, \citet{blh02}, \citet{lbs03}, and \cite{l04a}. The sample
does not include the recently published new Taurus brown dwarfs by \citet{gdm05}.

The millimeter observations were carried out in October and November 2005 using the IRAM
30\,m single-dish telescope on Pico Veleta/Spain, equipped with the bolometer MAMBO-2, which is
a 117-pixel array with a HPBW (half power beam width) of $\sim 11"$. The objects were centred
on the most sensitive pixel of the array, and since our target coordinates (taken from the
2MASS catalogue) are accurate within $\pm 1"$, we expect that this pixel contains the complete
flux from the targets. All observations were performed in ON/OFF mode for background subtraction,
using a wobble throw of 32". The integration times per subscan were either 10 or 60\,sec.
Some scans for target J043903+2544 were affected by additional noise due to a known 
acceleration problem of the bolometer (S. Leon, priv. comm.). Each target was observed on at least
two different nights, typically with on-source integration times of 20\,min per night, to
be able to identify problems with background subtraction, calibration, and increased noise.
During the observations, we aimed to reach comparable noise levels for all targets. As a
consequence, the total on-source times range between 40 and 90\,min (see Table \ref{targets}).

Data reduction was carried out using the MOPSIC pipeline provided by IRAM.
To check for inconsistencies, fluxes were measured for different scans separately. In general,
the flux levels obtained from different scans (but the same target) are comparable within
the noise limits. The final flux at 1.3\,mm was measured using all scans. These values are
listed in Table \ref{targets}. Typically, we reached noise levels of 0.7-0.8\,mJy.

\begin{figure}
\includegraphics[angle=-90,width=16cm]{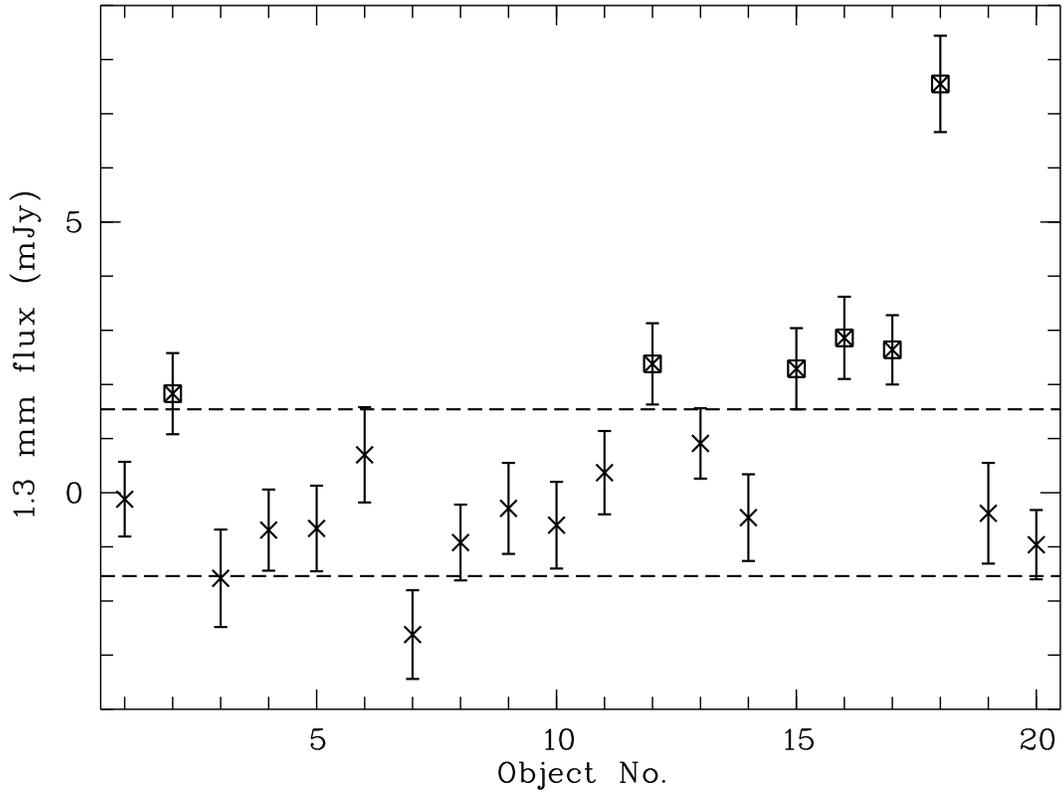}
\caption{1.3\,mm fluxes from IRAM bolometer observations for our 20 targets. The errorbars are 1$\sigma$
uncertainties; the dashed line indicates the 3$\sigma$ level for a {\it typical} noise level of 
0.77\,mJy. The six detections are marked with squares. The order of the datapoints corresponds
to the numbering in Table \ref{targets}. \label{f1}}
\end{figure}

In Fig. \ref{f1} we plot the 1.3\,mm fluxes obtained with IRAM for all targets, as a
function of their K-band magnitude. As can be seen from this figure, five objects show clear
positive detections with flux levels exceeding $3\sigma$. A sixth object (KPNOTau2) has a
$2.5\sigma$ detection. These six sources will be called 'detections' in the remainder of 
the paper. For object KPNOTau9, we obtain a 3$\sigma$ {\it negative} flux level, which is 
most likely related to excessive background emission. Excluding this object and the six 
detections, all remaining measurements scatter around zero.

The example of KPNOTau9 highlights that the derived fluxes may be affected by imperfect
background subtraction caused by other sources in the neighbourhood of our targets or inhomogenities 
in the Taurus cloud itself. To address this problem, we examined the bolometer maps for all 
our targets. Since the bolometer wobble throw was 32", only sources within this radius are likely
to contaminate the measurement significantly. Only one of our targets (J041411+2811) has a 
$>2\sigma$ detection within this distance, which is, however, only visible in parts of our scans.
Since our flux measurement is consistent in all scans, we consider it to be reliable. In a wider
radius of 100", two of our targets, KPNOTau9 and 12, have $>3\sigma$ neighbour sources. 
Both targets exhibit negative flux levels, in the case of KPNOTau9 on a 3$\sigma$ level, which
might be due to the emission from the nearby source. All other objects have maps without 
contaminating sources.

As a complementary test, we searched the IRAS point source catalogue for sources in a 100" 
circle around our targets. It turned out that only two objects -- KPNOTau9 and J041411+2811 -- 
have an IRAS neighbour. In the case of KPNOTau9, this neighbour has a 100$\mu m$ flux
of 17\,Jy and is located at a distance of 91", and thus possibly could contaminate the background 
measurement for the brown dwarf. We therefore attribute the negative flux level for KPNOTau9 to 
improper background subtraction. The IRAS neighbour of J041411+2811 probably has no significant 
influence on our 1.3\,mm flux measurement, as argued above. We examined the 100\,$\mu m$ IRAS 
images for all our sources, and found that the flux level in the region which affects the 
background subtraction, is more or less constant. These results confirm that our objects, with the 
exception of KPNOTau9, are fairly isolated and in regions without strong background inhomogenities.

An alternative assessment of the reliability of our mm fluxes can be made based on our
measurements itself. If the fluxes are pure noise (i.e. no significant emission from the target
and not affected by improper background subtraction), we expect them to scatter around zero, 
with Gaussian distribution. It is obvious that the complete sample is not consistent with
Gaussian noise, because we have six objects with $>3\sigma$  fluxes, whereas we expect zero.
After excluding all $>3\sigma$ detections (positive and negative), we expect 9.2 (66\%) to 
have flux levels within the 1$\sigma$ and 13.3 (95\%) within the 2$\sigma$ uncertainties. The 
actual numbers for our sample are 8 and 13, respectively. 

To verify if the average flux after excluding $>3\sigma$ measurements ($-0.2$\,mJy) is consistent 
with the expected zero value, we carried out Monte Carlo simulations: Assuming pure Gaussian 
noise with $\sigma = 0.78$ (the average uncertainty) and zero average, we generated 14 
datapoints and computed the average. From 10000 test runs, a substantial fraction of 15.6\% 
resulted in an average $\le-0.20$\,mJy. These two simple tests show that the scatter in Fig. 
\ref{f1} is fully consistent with Gaussian noise plus an excessive number of $>3\sigma$ 
outliers, confirming again that the quoted fluxes are (with the exception of KPNOTau9) most 
likely not severely affected by imperfect background subtraction. We therefore conclude that the 
fluxes for the positive $3\sigma$ detections are related to our substellar targets. Finally we 
note that two of our objects -- CFHTBDTau1 and 4 -- have already been observed with the same 
instrumentation \citep{kap03}; their 1.3\,mm upper limit (for CFHTBDTau1) and 
flux (for CFHTBDTau4) are completely consistent with our values.

We aimed to complement our 1.3\,mm fluxes with near-infrared and mid-infrared data to be able
to constrain the spectral energy distribution (SED) for the sources. This is particularly interesting
for the six detections, because it allows us to compare with disk models (see Sect. \ref{models}).
All detections except KPNOTau2 have been observed with the IRAC and MIPS instruments on board the 
Spitzer Space Telescope as part of the Spitzer GO program 3584 ('A Spitzer Imaging Survey of the 
Entire Taurus Molecular Cloud', PI D. Padgett). From these Spitzer images, we derived mid-infrared 
fluxes for these five sources and for the comparison object CFHTBDT2 (which has no mm detection). 
IRAF/daophot was used for aperture photometry. The apertures were chosen to avoid source confusion
and to optimize signal-to-noise (5 pixels for IRAC, 8 pixels for MIPS). Measured fluxes were converted 
to absolute fluxes using the aperture corrections given in the Data Handbooks for IRAC and MIPS. 
Near-infrared and optical photometry for the same six objects was taken from 2MASS and the 
surveys by \citet{mdm01} and \citet{l04a}. These magnitudes were converted to fluxes using the 
zeropoints given by \citet{s96}. For CFHTBDT4, we added to the SED the datapoints listed in Table 
1 of \citet{pah03}. These SED data will be used in Sect. \ref{models} to constrain disk properties 
of our targets. 

Since binarity might be an important factor for the disk properties (see Sect. \ref{dm3}), 
we searched for archived high-resolution images of our targets. Twelve of our 20 sources have 
been observed in the framework of the HST program no. 9853 (PI R. White) in deep exposures with 
three different filters. The reduced images are publicly available, and we used them to check 
for companions (Sect. \ref{dm3}).

\section{Disk masses}
\label{diskmass}

\subsection{Transforming fluxes to disk masses}
\label{dm1}

As outlined in Sect. \ref{obs}, we are confident that the obtained 1.3\,mm fluxes are related
to the observed substellar objects in Taurus. If this is the case, the mm flux is 
due to optically thin emission from circum-sub-stellar dust, which is directly proportional
to the dust mass \citep[see][]{bsc90}. The following equation relates the mm
flux $S_\nu$ to the dust mass $M_\mathrm{D}$:

\begin{equation}
M_\mathrm{D} = \frac{S_\nu D^2}{B_\nu(T_\mathrm{D})\kappa_\nu}
\end{equation}

\begin{figure}
\includegraphics[angle=-90,width=16cm]{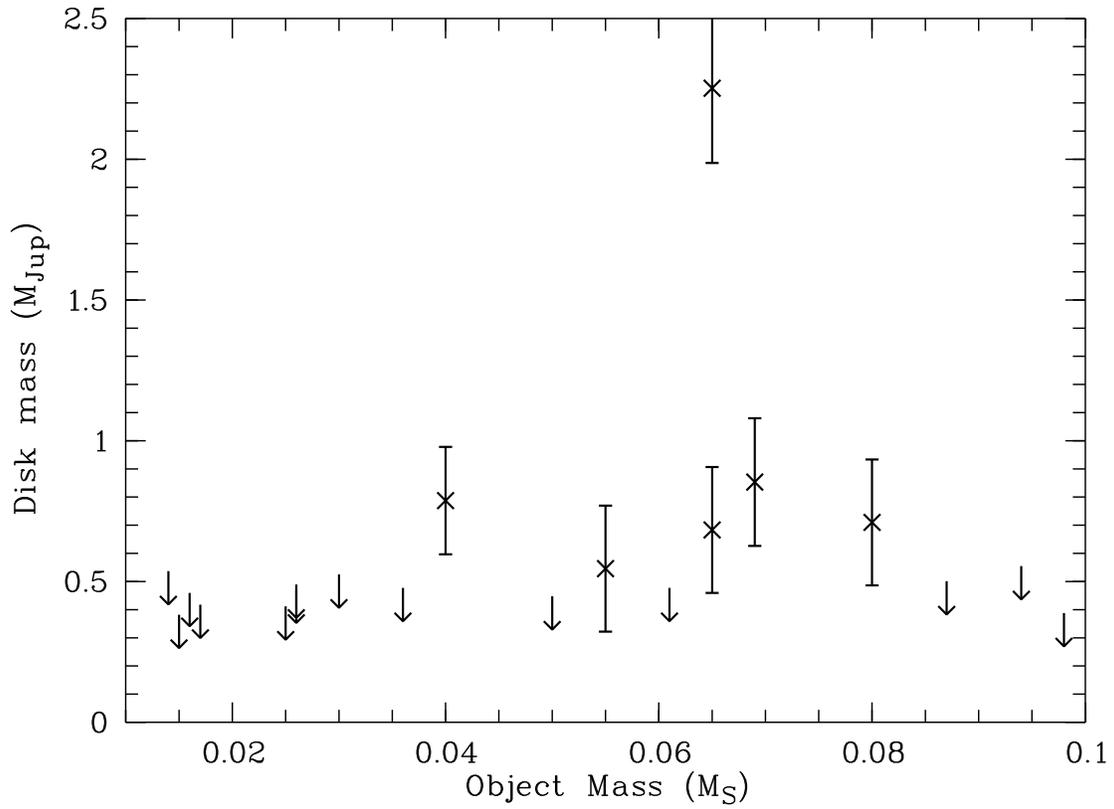}
\caption{Disk masses vs. object masses for our 20 targets. The errorbars are computed from the 
$1\sigma$ uncertainties for the fluxes and do not take into account the mostly systematical effects
of uncertainties in dust opacity, dust temperature, and distance. $2\sigma$ upper limits are shown for
objects without significant mm emission. Objects with very similar masses have been separated on
the x-axis for clarity. \label{f2}}
\end{figure}

The distance $D$ of our targets is known: All objects are spectroscopically
confirmed young members of the Taurus star forming region, for which a Hipparcos based distance
estimate of $142\pm 14$\,pc has been derived \citep{wbk98}. $B_\nu$ is the blackbody flux 
for the temperature of the dust $T_\mathrm{D}$. The plausible range for $T_\mathrm{D}$ 
is 10-20\,K, so we assume 15\,K here, consistent with, e.g. \citet{kap03} and 
\citet{ser00}. The dust opacity $\kappa_\nu$ is highly uncertain and not very well
constrained in the literature. However, when converting submm/mm fluxes to disk masses,
most groups use values between 1 to 3\,cm$^2$\,g$^{-1}$ at 1.3\,mm, in agreement with the
recommendations of \citet{oh94}. To be consistent with the literature values for disk masses 
for stars, we thus adopt a dust opacity of 2.0\,cm$^2$\,g$^{-1}$, as it has been used for
example by \citet{ser00,nby98, bsc90}. To convert from dust to disk masses we assumed a dust 
to gas ratio of 1:100, as generally adopted in the literature. The derived disk masses (or 
$2\sigma$ upper limits) using these parameters are listed in Table \ref{targets}, and plotted 
in Fig. \ref{f2}. For our six detections, the disk masses range from 0.55 to 2.55$\,M_\mathrm{Jup}$.

Dust temperature, opacity, and (to a minor degree) distance are the main sources of uncertainty 
in this calculation, and usually lead to large uncertainties in the disk masses. To assess the 
errors in our mass estimates, we carried out Monte Carlo simulations: We generated random numbers 
for $T_\mathrm{D}$, $\kappa_\nu$, and $D$ in the ranges given above, and computed 
disk masses for our six detections. The resulting range of likely disk masses is 
given in Table \ref{targets}. Please note that most of the uncertainty connected with
these three parameters is systematic and will affect all results in a similar way. The 
errorbars in Fig. \ref{f2} reflect only the 1$\sigma$ measurement uncertainty of the mm 
fluxes.

\subsection{Comparison with published disk masses}
\label{dm2}

Disk masses have been determined for large samples of T Tauri stars and for some
Herbig Ae/Be stars in star forming regions, but only for two brown dwarfs \citep{kap03}. 
This study provides the first large sample of disk masses for substellar objects. By comparing 
with literature results for higher mass stars, it allows us to study disk mass as a function of 
object mass over a mass range of more than three orders of magnitude -- from 0.02 to 3$\,M_{\odot}$. 
As outlined in Sect. \ref{intro}, such a comparison is an important tool to probe brown dwarf 
formation scenarios: If brown dwarfs are ejected stellar embryos, as predicted by a main class 
of formation models (see Sect. \ref{intro}), we expect disk truncation and thus reduced disk 
masses. This may lead to a break or a trend towards lower disk masses in the substellar regime. 

Ideally, one has to compare disk masses for coeval objects, to exclude influence of an age
spread. The disk masses of T Tauri stars show only little dependence on age for 
objects with ages between 1 and 5\,Myr \citep[see e.g.][]{ncz97,aw05}.
The disk masses decrease significantly for stars with ages $>5\,$Myr, but such 
objects are rare in Taurus. Most of the young stellar objects in Taurus are known to have 
ages between 1-3\,Myr \citep{l04a}. Thus, for our purposes, we consider the 
substellar Taurus population to be coeval. It is thus legitimate to compare our brown dwarf 
disk masses with those of T Tauri stars in star forming regions with similar ages. The most 
appropriate way to do this is to compare the {\it ratio} of disk mass to object mass. 

We estimate masses for our brown dwarf targets by converting their spectral types 
to effective temperatures using the scale by \citet{lsm03} and comparing these 
temperatures with the most recent evolutionary tracks by \citet{bcb03} assuming an 
age of 2\,Myr. The derived object masses are listed in Table \ref{targets}. Although the 
uncertainties in the conversion from spectral types to masses are considerable, mainly due 
the evolutionary tracks at young ages \citep[see][]{bca02}, all our targets 
are likely to have masses between 0.01 and 0.1$\,M_{\odot}$. We note that the model 
uncertainties lead to systematic errors, thus the relative masses in our sample are more 
reliable. The same procedure was applied to derive masses for the young brown dwarfs and very
low mass stars observed by \citet{kap03}, which are included in the following analysis, in
cases that we did not observe in our survey.

As comparison samples for higher mass stars, we used the results from \citet{ob95},
\citet{ncz97}, \citet{nby98}, and \citet{ngm00}. All these papers contain lists with 
object masses and ages as well as disk masses determined either from SED modeling or mm measurements 
for objects mainly belonging to star forming regions in Taurus, $\rho$\,Oph, and Lupus. Again, all 
object masses should be considered as rough estimates, but for our purposes even uncertainties of 
100\% are tolerable. To avoid being biased by an age spread (see above), we used only the subsample 
of stars with ages between 1 and 5\,Myr, where disk masses are known not to show a significant
trend with age. All comparison stars can thus be considered to be coeval with our brown dwarfs. 
The final sample of comparison sources comprises of 52 objects with masses between 0.03
and 2.7$\,M_{\odot}$, among them 25 upper and 2 lower limits. Their ratios of disk and object mass 
are plotted in Fig. \ref{f3} along with our values for Taurus brown dwarfs.

\subsection{Discussion}
\label{dm3}

Fig. \ref{f3} does not reveal a significant change of disk to object mass ratio with
object mass. Specifically, there is no obvious difference in the very low mass regime, 
i.e., among the lowest mass stars and brown dwarfs, where ejection might 
become important \citep{gww04}. The average mass ratio (for the detections) is 1.9\% in the 
stellar and 2.6\% in the substellar regime, thus comparable given the large scatter and 
uncertainties. There is no hint of an underabundancy of 'massive' disks among brown dwarfs: 
$41\pm 14$\% (9 out of 22) detected disks around stars have ratios $>2$\%, whereas there are 
two such objects among brown dwarfs -- 2 out of 6, i.e. $33\pm 23$\%. If we consider all objects, 
there are three stars ($6 \pm 3$\%) with ratios $>5$\%, and one brown dwarf, i.e. $11\pm 11$\% 
(IC348-613 from \citet{kap03}). Here we neglect upper limits close to 5\%, because these objects 
are unlikely to have disk masses $>5$\%. Taking into account the large uncertainties in disk and 
object masses, there is no statistical basis to claim that brown dwarfs lack (relative) massive 
disks. 

There is still the possibility that such an effect is hidden in the non-detections. 
The average disk to object mass ratio is $1.9$\% in the stellar regime, which is an upper limit, 
because it does not take into account non-detections. Assuming no trend with mass, we expect half
of the brown dwarfs to have values below 1.9\%. For nine of our brown dwarfs we can rule out
that they have ratios $>1.9$\%, and for two of them we know that the ratio is $>1.9$\%. Thus,
if most of the upper limits with ratios $>1.9$\% are in fact values lower than 1.9\%, there
would be reason to believe that brown dwarfs have more often very little circumstellar
material than stars, which would in turn be indirect evidence for truncated disks. Based 
on the available observational data, this possibility cannot be definitely excluded. The next 
generation of submm telescopes will hopefully provide the means to verify this hypothesis.

Recapitulating, we do not see any overall trend of relative disk mass with object mass.
The dominant feature seen in Fig. \ref{f3} is a large scatter over the entire mass range, which 
appears to be more significant than any possible mass dependency. Previous studies of this problem
are inconclusive: \citet{aw05} find that (absolute) disk masses in Taurus-Auriga 
scatter over three orders of magnitude and do not show any trend with stellar masses (which range 
from 0.1 to 2.5$\,M\,_{\odot}$ in their sample). The same result has been obtained by \citet{ms00}, 
who include datapoints for Herbig Ae stars with masses between 1 and 
4$\,M_{\odot}$. If, however, the range of disk masses is constant over such a large mass range, 
this implies an increase of the relative disk mass with decreasing stellar mass by about one 
order of magnitude. This is in agreement with the studies of \citet{ncz97} and \citet{nby98}, who 
find a weak correlation of relative disk mass and stellar mass in Lupus and $\rho$\,Oph, in the 
sense that stars with low masses ($<0.7\,M_{\odot}$) tend to have more massive disks.  

On the other hand, the review paper by \citet{ngm00} does not report any evidence for such
a trend. They compile disk masses for T Tauri and Herbig Ae/Be stars, and find
a positive correlation of (absolute) disk mass with stellar mass. Converted to relative
disk mass, the correlation disappears. Consequently, they claim that the ratio of disk to 
stellar mass is roughly constant in the mass range from 4 to 0.3$\,M_{\odot}$. This is confirmed
by our Fig. \ref{f3}, which does not show any clear sign of a trend. In any case, if there is
a correlation, it is much weaker than the scatter in disk masses. 

One reason for the discrepancy in the literature results might be an age spread or environmental
effects on disk masses. By throwing together measurements for objects with ages varying from
0.1 to 10\,Myr, as done in most literature studies, possible trends with object mass might
be diluted, which was the reason for us to include only objects with ages between 1 and 5\,Myr in our
analysis. Strictly speaking, one has to separate age and mass effects, but this leads to very
small samples. 

An important reason for the consistently large scatter in all previous studies of disk masses as well
as in our own data might be binarity. It has long been known that the existence of a companion
and the binary separation clearly affects the submm properties of T Tauri stars \citep{ob95,nby98}. 
Separations smaller than 50-100\,AU appear to inhibit the submm flux \citep{ob95,jmf94,jmf96}, 
presumably by truncating the outer disk. For very close companions ($<1$\,AU), \citet{jm97} found 
evidence for cleared out inner regions, which strongly affects the mid-infrared, but not the mm fluxes. 
The recent analysis of \citet{aw05} finds that the presence of a companion with separation 1-100\,AU 
tends to decrease the mass of circumstellar material, although they still have detectable disks.

If and how this trend continues in the substellar regime is unknown. We used deep images from the
HST, which are available for 12 of our sources (all KPNO and CFHT objects), to assess the binarity
in our brown dwarf sample by careful visual inspection. Only one of the 12 objects appears to have a 
companion within $3\farcs5$, corresponding to 500\,AU in Taurus. KPNOTau9, the only object with 
an obvious 
companion, shows a neighbour at $\sim 250$\,AU. We did photometry on images in the HST filters corresponding 
to z' and i band, and found that the magnitude difference is $5.0$\,mag in the red, but only $4.6$\,mag in 
the blue filter. If it is a physical companion or at least a member of the Taurus star forming region, we 
expect it to be significantly redder than the primary. This is not the case, so we conclude that we are
seeing a background object. To verify the lower detection limit in separation, we additionally checked three 
known binary brown dwarfs in Upper Scorpius, which have been detected with the same dataset \citep{kwh05}. 
Two companions with separations of $0\farcs12$ and $0\farcs07$ are clearly seen by visual inspection, whereas 
the third one with separation $0\farcs03$ is not obvious, and can only be recovered with PSF fitting. 
Thus, a conservative estimate for our inner detection limit is $0\farcs07$ or $\sim 10$\,AU. 

We conclude that none of the twelve objects observed with HST is a binary with separation 
between 10 and 500\,AU, and thus the frequency of companions in this separation range appears 
to be quite low. From our analysis, we derive an upper limit of 22\% (with 95\% confidence) 
for separations between 10 and 500\,AU. This is in line with high-resolution imaging surveys of 
field brown dwarfs that find very few companions with separations $>20$\,AU \citep{bbm03,mbb03},
although recently \citet{bmb06} found a possible population of wide companions in the Upper 
Scorpius star forming region. At least for our targets in Taurus, we expect no significant impact from 
wide companions on the disk properties in the substellar regime. Close companions with separations 
$<10$\,AU may still exist around our objects and affect their disks (although probably not the disk 
mass). Recent results indicate a spectroscopic binary frequency of 11\% in the very low mass regime 
for field objects \citep{br06}.  
In the absence of comprehensive binary studies for brown dwarfs with disks, it is difficult  
to disentangle the effects of binarity and the formation process on the properties of the disk.

Pondering all these observational results, we arrive at three conclusions: a) The dominant
effect in the existing database of disk masses is a large scatter over at least two orders 
of magnitude, independent of object mass, which might be due to binarity. b) We do not see 
any trend of relative disk mass with object mass, in contrast to several literature studies.
c) Based on the existing mm data, there is no clear break at or around the substellar limit, 
indicating no significant change in the disk properties due to, for example, an ejection 
process.

\section{Modeling the SEDs of brown dwarf disks}
\label{models}

\subsection{Model description}

We use Monte Carlo radiation transfer codes to generate model SEDs for dusty disks 
irradiated by brown dwarfs.  Given the apparent low accretion rate of brown dwarfs 
\citep{mhc03} for all models we assume disk heating is dominated by radiation from 
the brown dwarf. In all our models presented below, for the incident stellar spectra 
we use NextGen model atmospheres \citep{aha01,hab99} with $\log g = 4.0$.  
The combination of mm and IR data allows us to investigate disk masses, sizes, 
and structures. For the disk structure we explore the following two scenarios: disks in 
vertical hydrostatic equilibrium with dust and gas well mixed \citep{dcc98,wwl04}; and 
geometrically flatter disks where the dust and gas are not coupled and grain growth and 
sedimentation of large grains towards the midplane has occured \citep[e.g.][]{mn95,dd04,dch06}. 
For the disks in vertical hydrostatic equilibrium we use the models described by \citet{wwl04} 
where the radial gradient of the disk surface density is described by a power law, 
$\Sigma(R)\sim R^{p}$. Disk surface density gradients are usually in the range 
$-2\le p \le -1$, with $p=-3/2$ the value quoted for the minimum mass solar nebula 
\citep{h81} and $p\approx -1$ found for irradiated steady accretion disks \citep{dcc98}. 
In all our models below we use $p=-1$. 

Hydrostatic equilibrium disks around brown dwarfs are highly flared due to the lower 
gravity of the central star \citep{wwl04} and as will be demonstrated below, such models 
do not provide good fits to our Taurus brown dwarf disks.  Therefore, to investigate 
deviations from hydrostatic equilibrium (e.g., dust settling and geometrically flatter 
disks) we adopt the following parameterization for the two dimensional density structure 
of the disks \citep{ss73}
\begin{equation}
\rho=\rho_0 \left ({R_\star\over{\varpi}}\right )^{\alpha}
\exp{ -{1\over 2} [z/h( \varpi )]^2  }
\; ,
\end{equation}
where $\varpi$ is the radial coordinate in the disk mid-plane and the scaleheight increases with 
radius, $h=h_0\left ( {\varpi /{R_\star}} \right )^\beta$. We vary the degree of flaring within 
the geometric disk models by adjusting the values of $\beta$ and $h_0$. As with the hydrostatic 
models, we assume the surface density exponent $p=\alpha-\beta-2 = -1$. This allows us to 
investigate to a first approximation the degree to which the disk structure deviates from the 
vertical hydrostatic equilibrium case. Our approach of using two grain populations 
is similar to other recent work on SED modeling \citep{dd04,dch06}.

As some brown dwarf disks show evidence for silicate features in their spectra \citep[e.g.][]{pah03}, 
we have modified our code to include multiple grain size distributions that have different spatial 
distributions \citep[e.g.][]{w03,cbm04}. The process of dust growth and sedimentation is thought 
to result in small grains remaining coupled to the gas and larger grains settling towards the disk 
midplane \citep[e.g.]{jld02,dd04,dch06}. Therefore for this present investigation we adopt two 
different grain size distributions and assign them different scaleheights.  We assume that small 
grains with an interstellar-like size distribution \citep{kmh94} have a larger scaleheight than 
the larger grain size distribution we have used previously to model SEDs of disks around Classical 
T~Tauri stars \citep[e.g.][]{wwb02,otw05,sws03}. This dust model includes silicates and carbonaceous 
grains using solar abundance constraints.  The grain size distribution is a power law with an 
exponential decay for particles with sizes above $50\mu$m and a formal maximum grain size of 1~mm 
\citep[see desription of grain model in][]{wwb02}. Although our large grain model has been successful in 
fitting SEDs of other disk systems, we do not claim our particular grain model represents {\it the} 
dust size distribution in all disks and note that it is the product of opacity and mass that may be 
determined from SED fitting of long wavelength data.  

For simplicity and in the absence of a detailed model for dust growth and settling, we use eqn. 2 
to describe the density structure of the two grain models. Both components have the same surface 
density distribution ($\alpha$ and $\beta$) and we vary $h_0$ for each component to simulate the 
different scaleheights arising from settling of the larger grains. The disk mass is dominated by 
the large grain component and we add a small amount of ISM-like grains with a larger scaleheight 
to fit the silicate features present in the SED of CHFTBDT4. In our parameterization we assign 
the mass $M_d$ to the large grains and the mass $f_{\rm ISM} M_d$ to the ISM-like grains.  The total 
disk mass is then $(1+f_{\rm ISM})M_d$, where this is the mass of dust plus gas with an assumed gas 
to dust ratio of 100.

For all models we assume that dust in regions close to the star is destroyed if temperatures 
rise above 1600K \citep{dgt96}. This condition provides a minimum inner dust radius of typically 
$\sim 6R_\star$. Any remaining gas within this gap we assume to be optically thin and therefore 
we effectively have an opacity gap in the disk \citep{la92}. All models are subject to reddening 
using the extinction curve for interstellar grains from \citet{kmh94}.

\subsection{Results: Disk structure, masses and radii}

As discussed in many papers, there are lots of degeneracies in fitting SED data: disk structure, 
scaleheights, radii, surface density, dust properties \citep[e.g.][]{cjc01}.  
Robust determinations of the disk structure require modeling of multiwavelength imaging as well as 
spectroscopy and so far only a few sources have sufficient data to allow such a study (e.g., see the 
combined scattered light and SED modeling of GM~Aur in \citet{sws03}). The current data available 
on brown dwarf disks is limited to SED data with large gaps in wavelength coverage --- the sources we 
are modeling have no data in the range $25\mu{\rm m}\le\lambda\le1300\mu{\rm m}$, which is a crucial 
regime for determining disk structure.  Therefore, the fits presented are ``by eye" and we have not 
attempted any sophisticated least squares fitting of the data.  We are not claiming the model 
fits presented are unique solutions for the disk structure, but they do allow us to address the 
following questions: what is the minimum disk radius, is there evidence for dust 
growth and sedimentation (i.e., flatter disks), and is there evidence that the disk structure 
significantly deviates from vertical hydrostatic equilibrium?

\begin{figure}
\center
\includegraphics[angle=0,width=12cm]{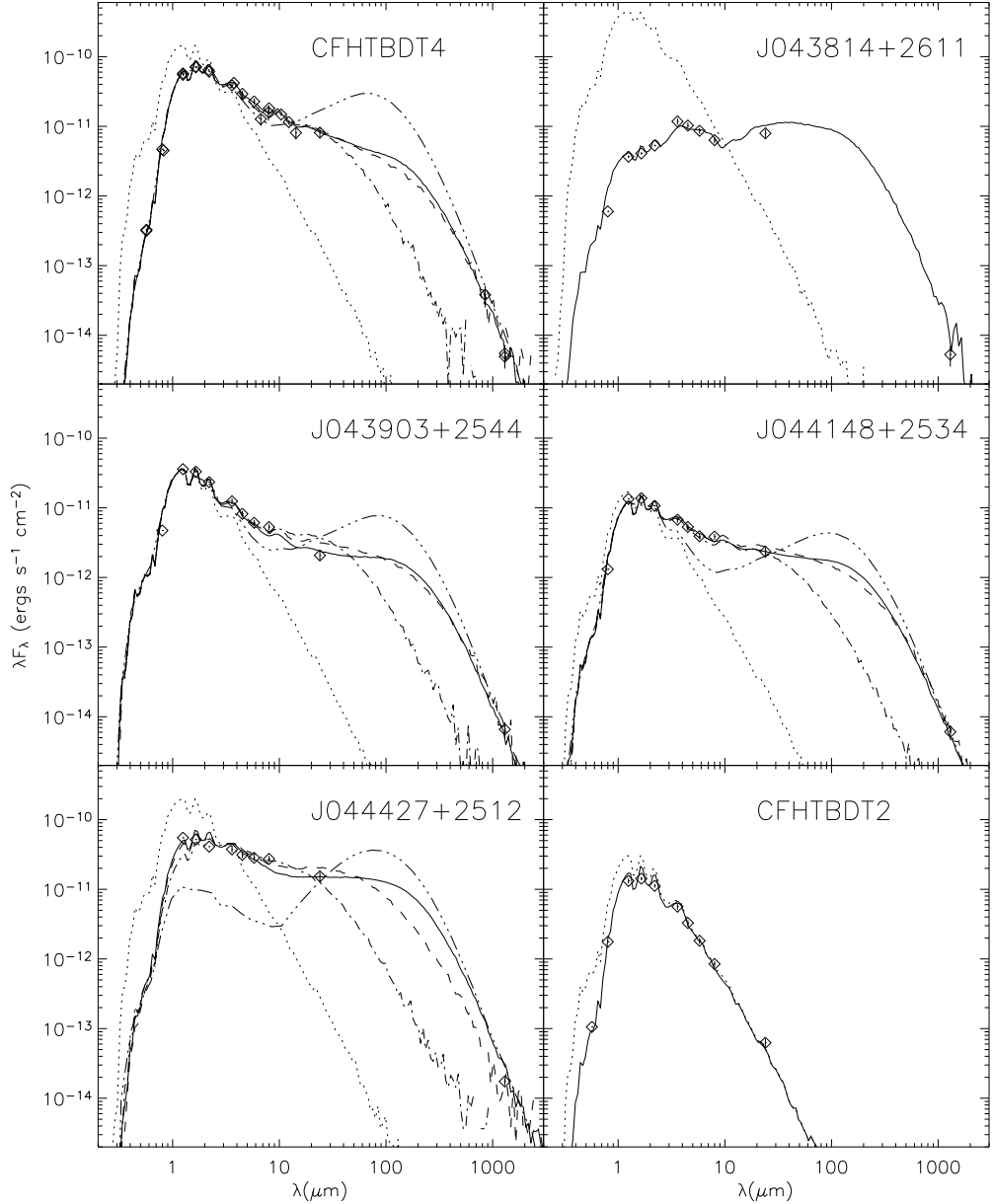}
\vspace{0.8cm}
\caption{SED data and model fits for the six sources discussed in the text. Most panels show
five curves: input stellar spectrum (dots), best fitting disk model using two dust components with 
$R_D=100$~AU (solid), two-component disk model with $R_D = 10$~AU (dashed), 
two component disk model with $R_D = 1$~AU (dot-dash), 
and a hydrostatic model with the same total mass as the best fitting disk model and radius $R_D = 100$~AU 
(dash-triple dot).  For the almost edge-on system J043814+2611, we only show the best fit two-component 
dust model with $R_D = 100$~AU.  For the moderately inclined system J044427+2512 the hydrostatic disk 
structure is more vertically extended than our two-component model and as such obscures the star at 
optical and near-IR wavelengths. For CFHTBDT2 the data are consistent with no disk and only the stellar 
atmosphere (dots) and reddened stellar atmosphere (solid) are shown.
\label{f4}}
\end{figure}

The results of our modeling are displayed in Fig. \ref{f4} and the model parameters used for each source 
are displayed in Table \ref{modelparams}. Each panel in Fig. \ref{f4} shows the data and five curves: the 
input stellar atmosphere model, best fitting disk model using two dust components with 
$R_D=100$~AU, two-component disk model with $R_D = 10$~AU, two component disk model with $R_D = 1$~AU, 
and a hydrostatic model with the same total mass as the best fitting disk model and radius $R_D = 100$~AU.  
As one of our goals is to determine whether the data will allow us to place constraints on the 
{\it minimum} disk radius, we have not explored disk models with radii greater than 100~AU.  

Immediately we see that the hydrostatic disk models produce too much mid-infrared emission. 
This is due to the large scaleheights in hydrostatic disks around low mass objects \citep{wwl04}.  
Better fits to the SEDs are provided by our two-component disk models where the large grains have 
a smaller scaleheight than the small grain population.  This result is consistent with previous 
investigations \citep[e.g.][]{tno02,pah03} and implies that brown dwarf disks do not have dust 
and gas well mixed in vertical hydrostatic equilibrium. However, we also see that in general the 
hydrostatic models underpredict the near-IR excess emission. Compared to our two-component power 
law disk models, this implies that the inner disk regions have scaleheights larger than hydrostatic 
models and the outer disk regions are less flared with scaleheights that are smaller than the 
corresponding hydrostatic solution. The cause of this ``super-hydrostatic'' scaleheight in the 
inner disk regions clearly cannot be due to radiation from the central star; otherwise the 
hydrostatic models would fit the near-IR flux levels.  

This ``super hydrostatic'' effect in the inner disk is also present in models of Herbig Ae 
disks \citep[][C.~Dullemond, private communication]{vij06}. For these systems SED models also 
require the inner disk scaleheight to be larger than that for a disk in vertical hydrostatic 
equilibrium. At present we do not know the reason for this ``super-hydrostatic'' scaleheight 
effect. One possibility is reprocessing of stellar photons in low density material close to the 
star such as a disk wind. It may also be that some other heating mechanism, e.g. chromospheric
or coronal emission, is responsible for the observed near-IR excess. Finally, close companions
may be able to interact with the disk and transfer energy to the inner parts of the disk. More 
detailed dynamical models of disk structure and evolution are clearly required to further 
investigate this effect.  

The disk masses derived from our radiation transfer simulations are consistent with the
disk mass ranges estimated from the mm fluxes alone using the simple formula of equ. (1), 
in particular if we take into account the different mm opacity in our 
adopted dust model, $10\, {\rm cm}^2\,{\rm g}^{-1}$ compared to 
$2\, {\rm cm}^2\,{\rm g}^{-1}$ used in Sect. \ref{dm1}. Four of our sources may be modeled with 
disk masses $M_D=4\times 10^{-4}M_\odot$ and one has a larger mm flux and a correspondingly higher 
$M_D=1.2\times 10^{-3}M_\odot$.  Note that these masses are the total mass of dust plus gas that 
contributes to the observed SED and does not include a sizeable mass of dust with grain sizes above a 
few hundred microns or very large objects (rocks, boulders, planets) that may be present in the disk. 
As such, the derived masses are the minimum circumstellar mass for the dust model we have used.

Our models show that with current data we cannot discriminate among models that have radii greater 
than $R_D\sim 10$~AU.  Smaller disks cannot reproduce the mm data, even if they are very massive.  
This is demonstrated by the $R_d=1$~AU models which have $M_d=1\,M_\odot$.  These disks cannot reproduce 
the mm flux levels because there is not enough material at cool enough temperatures to provide emission 
at long wavelengths.  Therefore, the mm flux levels observed mean that the disks must have radii greater 
than 10~AU.  

We now give some brief comments on the individual sources in Fig. \ref{f4}.

{\it CFHTBDT4:}  This source was previously modeled by \citet{pah03}. Our modeling agrees well 
with theirs for the luminosity and interstellar extinction, though our use of a model atmosphere 
instead of a blackbody provides a better match to the shortest wavelength fluxes.  We achieve a good 
match to the data around the 10$\mu{\rm m}$ silicate feature using ISM grains in an extended layer with 
a scaleheight twice that of the larger grains in our models. It is very difficult to discriminate 
the $R_D = 100$~AU and $R_D=10$~AU models, even at far-IR wavelengths \citep[see also][]{bsc90,cjc01}. 
Disk radii smaller than 10~AU cannot reproduce the SED, even if the disk is very massive.

{\it J043814+2611:}  This source is under-luminous at near-IR wavelengths and was suspected of being 
a disk viewed close to edge-on \citep{l04a}. Our models confirm this as we find a viewing 
angle $i=80^\circ$ for the star plus disk system.  The luminosity of the central source in our models 
is larger than that estimated from spectral typing.  This is not surprising because disks viewed 
edge-on are seen only via scattered starlight so photometric spectral typing is prone to error. In 
addition Walker et al. (2004) pointed out that edge-on Classical T~Tauri stars may have colors 
resembling less inclined brown dwarf plus disk systems.  Due to the uncertainty in the spectral 
type we have not attempted any other models for this source (small disk radii or hydrostatic disks).

{\it J043903+2544:} The hydrostatic model for this source clearly produces too much emission at 
$24\mu$m.  The ``super hydrostatic" effect for the inner disk is seen by the poor match of the 
hydrostatic model at IRAC wavelengths.  The two-component disk model provides a much better match 
to the data.

{\it J044148+2534:} The hydrostatic model for this source does match the $24\mu$m data, but at 
IRAC wavelengths the ``super hydrostatic'' effect is very pronounced.

{\it J044427+2512:} We derive a viewing angle of around $i=63^\circ$ for this source and a 
luminosity that is larger than estimated from spectral typing. As with J043814+2611, inclination 
effects are likely contributing to confusion in the photometric estimate of the spectral type for 
this source.  The hydrostatic disk structure is more vertically extended than our two-component model 
and as such obscures the star at optical and near-IR wavelengths for $i>60^\circ$.

{\it CFHTBDT2:} This source exhibits no excess emission at infrared wavelengths and we do not 
detect it at 1.3\,mm.  The source is modeled with a brown dwarf model atmosphere with $T_\star=2700$~K, 
$R_\star = 0.61$, giving $L_\star = 0.018 L_\odot$.  

To summarize our SED modeling, we find that the sources detected at 1.3\,mm
have disks with masses in the range 0.4 to 1.2\,$M_J$ and outer radii greater than
10\,AU. In general we find the disks have scaleheights in the outer disks
that are smaller than would be expected for a disk in vertical hydrostatic
equilibrium around a brown dwarf, thus implying dust settling to the disk
midplane. However, the near-IR excess emission appears to require the inner 
disks to be more vertically extended than the hydrostatic solution. It is
unclear whether this is related to a true vertical thickening of the inner 
disk or the reprocessing of stellar radiation in low density material near 
the star such as a disk wind.

\section{Summary}
\label{sum}

This paper presents a study of brown dwarf disk properties based on the largest
sample of mm measurements for this object class so far, complemented by
mid-infrared datapoints from Spitzer. We used the fluxes at 1.3\,mm to constrain
disk masses and the SED to constrain disk radii and geometry by comparing with
SED models.

The masses of brown dwarf disks range from fractions of one Jupiter mass up
to a few Jupiter masses. The relative masses of brown dwarf disks are not 
significantly different from values measured for coeval stars. Most of them 
range between $\lesssim 1$ and 5\%, with a few outliers between 5 and 10\%. A 
substantial fraction of brown dwarf disks (at least 5 out of 20 in our sample) 
have radii larger than 10\,AU. Smaller disks do not have enough cool material to 
provide the emission observed at 1.3\,mm.

What are the implications of the results given above for our understanding of 
brown dwarf formation? As outlined in Sect. \ref{intro}, the two leading theories
for brown dwarf formation are collape of (isolated) cores ({\it in situ} formation)
and ejection from multiple systems. The latter process will affect disk properties
and probably leads to truncated disks. Our constraints for disk masses and radii of 
brown dwarfs do not provide any evidence for the existence of these truncated disks. 
Particularly there is no change of the relative disk mass around the substellar limit. 
Moreover, we find $>25$\% of our targets to have disk radii $>10$\,AU, whereas the
prediction for an ejection scenario is $\sim 5$\% for radii $>$10-20\,AU 
\citep{bbb02,bbb03}. Truncated disks are expected to evolve viscously to larger
radii after the ejection process \citep{bbb03}, but in this case they will have
very low masses. Since we have found disks with radii $>10$\,AU {\it and} relative 
masses comparable to those of stars, they are unlikely to be truncated.

Thus, in our survey brown dwarfs appear to harbour scaled down T Tauri 
disks. Thus, the disk properties found in this paper and in the literature 
\citep{pah03,kap03,mjn04} are completely consistent with a scenario where brown 
dwarfs form {\it in situ}, i.e. from isolated molecular cloud cores with very 
low masses. By applying Occam's Razor, we conclude that there is no need to invoke 
an ejection process. On the other hand, if ejection plays a role in brown dwarf 
formation, it is unlikely to be responsible for all objects.

However, although the interpretation of the available information on brown dwarf 
disks does not require an ejection mechanism, it cannot firmly rule out ejection either.
The high number of non-detections in our mm observations may indicate that we are 
simply not sensitive enough to find the signature of truncated disks. Thus, it may be 
that the effect of an ejection process is hidden in the non-detections in our sample. Also, 
the large number of free parameters in the SED models and the large gap in wavelength coverage 
in the observations still allows for a broad range of disk radii. On the other hand, the 
theoretical predictions for the outcome of close encounters in multiple stellar systems, 
which eventually would lead to truncated disks, are sparse and mostly not quantitative. It 
is clear that some ejected objects can retain a substantial fraction of circum(sub)stellar 
material, but masses and radii of the ejection-affected disks are poorly constrained. To 
distinguish between ejected brown dwarfs and {\it in situ} brown dwarfs by studying disk 
properties at ages of 1-5\,Myr additionally requires assumptions about the disk evolution 
on this timescale. 

Whereas the non-detections in our observations might hide the signature of ejection, 
it is unlikely that {\it all} disks in our sample and the previously studied brown dwarf 
disks have experienced an ejection. At least in a few cases, brown dwarf have disks with 
$>2$ Jupiter masses, which corresponds to 3-9\% of the mass of the central object. It is 
unlikely that these (relatively speaking) massive disks have been truncated and lost a 
significant fraction of their mass during an ejection process. The large fraction of objects 
with disk radii $>10$\,AU is also hardly consistent with formation only by ejection. This 
simply implies that at least a fraction of brown dwarfs forms in isolation, without violent 
event in their early evolution. More indications for this interpretation comes from the recent 
discoveries of a few very wide brown dwarf binary systems \citep{l04b,cld04}. Thus, for a 
subsample of brown dwarfs there is clear evidence for isolated formation. Future studies 
should aim to quantify the fractions of substellar objects formed via ejection and/or 
{\it in situ}. 

The derived disk masses and upper limits put constraints on possible planet formation scenarios
in brown dwarf disks. Since only a small fraction of brown dwarfs have disk masses larger than 
$>1\,M_\mathrm{Jup}$, it is unlikely that Jupiter-mass planets are frequent around brown dwarfs.
Here we assume that planets have not already formed at 1-5\,Myr, because our disk mass estimates
are only based on the dust. On the other hand, there is certainly enough circum-sub-stellar
material to form less massive planets. 

\begin{figure}
\includegraphics[angle=-90,width=16cm]{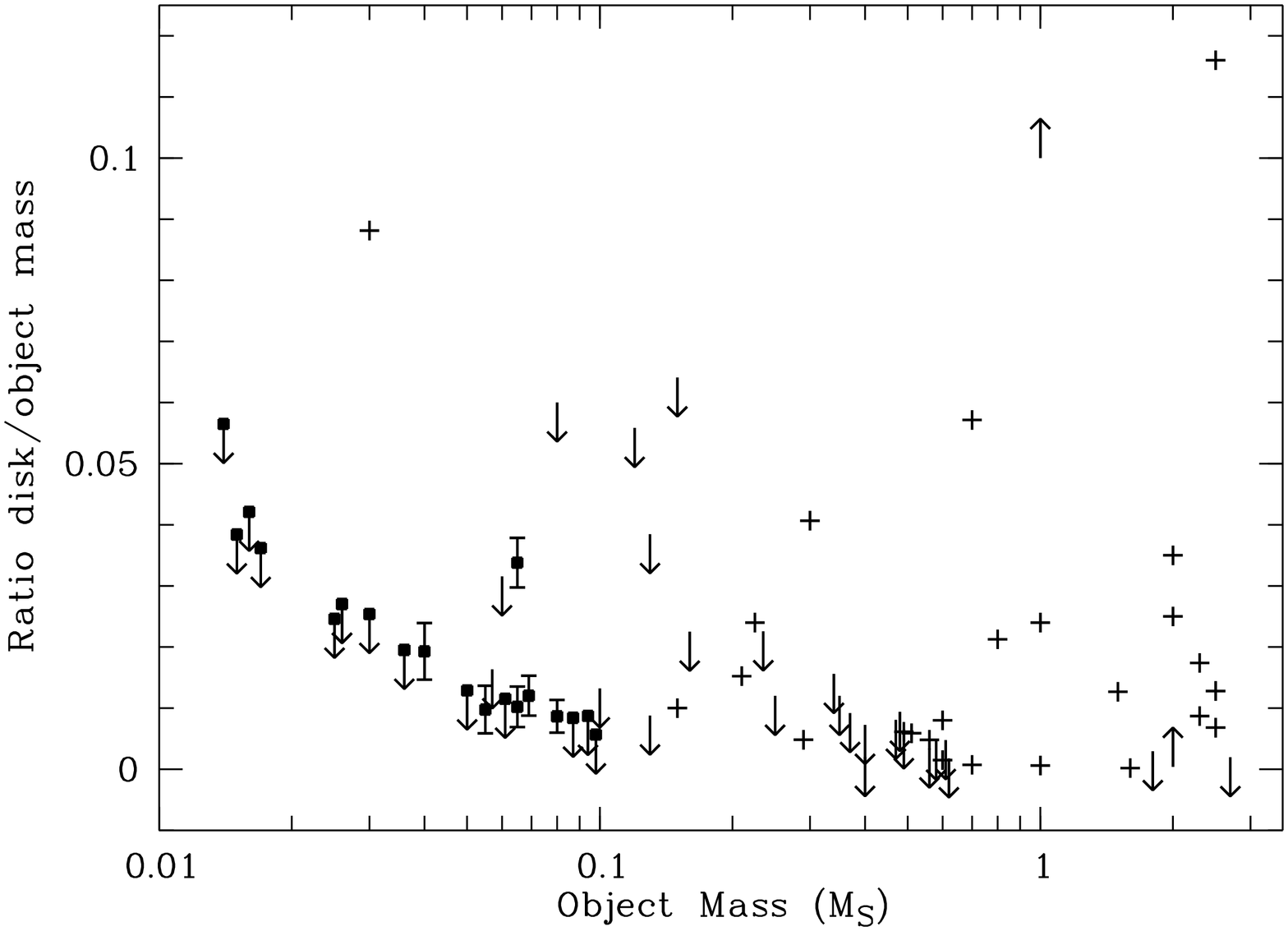}
\caption{Ratio disk mass to object mass vs. object mass. The plots includes our measurements for
brown dwarfs (marked with filled small squares) as well as data from \citet{kap03,ob95,nby98,ncz97,ngm00} 
({\bf +}). Arrows show upper or lower limits. For the Taurus brown dwarfs, upper limits are based on 
2$\sigma$ flux upper limits. Objects with very similar masses have been separated on the x-axis for clarity.
\label{f3}}
\end{figure}

Our mm observations for brown dwarfs extend the mass range of objects with constraints
on disk masses down to $\sim 0.02\,M_{\odot}$ (see Fig. \ref{f3}). In our age-spread corrected 
sample, we do not see any significant change of relative disk mass with object mass, as claimed 
previously in the literature. The dominant feature in this plot is a large scatter of at least 
two orders of magnitude, independent of the mass of the central object. Apparently, the disk masses 
only scale down with object mass. As outlined in Sect. \ref{diskmass}, binarity may play
a more important role than central object mass for the evolution of the disk.

\acknowledgments
It is a pleasure to acknowledge the support from the IRAM pool observing team, in particular
Stephane Leon. Christina Walker and Mark O'Sullivan kindly helped with the modeling section of 
this paper. We thank Mirza Ahmic for assisting us in the selection of Spitzer images for our 
targets. We are grateful to an anonymous referee and Matthew Bate for helpful suggestions that 
improved the paper. This work is based in part on observations made with the Spitzer Space 
Telescope and on data products from the Two Micron All Sky Survey. We make use of observations made 
with the NASA/ESA Hubble Space Telescope, obtained from the ESO/ST-ECF Science Archive Facility. 
This research was supported by an NSERC grant and University of Toronto startup funds to R.J. 

Facilities: \facility{IRAM, Spitzer, HST}

\clearpage

\begin{deluxetable}{llcrccc}
\tabletypesize{\scriptsize}
\tablecaption{Targets, fluxes, masses \label{targets}}
\tablewidth{0pt}
\tablehead{
\colhead{Full name} & \colhead{SpT} & \colhead{Int. time} &
\colhead{1.3\,mm flux} & \colhead{Disk mass\tablenotemark{a}} & \colhead{Object mass} & \colhead{disk 
mass range\tablenotemark{b}} \\
\colhead{} & \colhead{} & \colhead{(min)} & \colhead{(mJy)} & \colhead{($M_J$)} &
\colhead{($M_J$)} & \colhead{($M_J$)}}
\tablecolumns{8}
\startdata
KPNOTau1      & M8.5  & 62  & $-0.12 \pm 0.69$ & $<0.42$         & 0.026 & \\	                     
KPNOTau2      & M7.5  & 60  & $1.83 \pm 0.75$  & $0.55\pm 0.22$  & 0.052 & $0.3\ldots 1.6$\\         
KPNOTau4      & M9.5  & 55  & $-1.58 \pm 0.90$ & $<0.52$         & 0.014 &\\	     		     
KPNOTau5      & M7.5  & 52  & $-0.69 \pm 0.75$ & $<0.44$         & 0.052 &\\	     		     
KPNOTau6      & M8.5  & 86  & $-0.66 \pm 0.79$ & $<0.47$         & 0.026 &\\	     		     
KPNOTau7      & M8.3  & 51  & $0.70 \pm 0.88$  & $<0.52$         & 0.030 &\\	     		     
KPNOTau9      & M8.5  & 48  & $-2.62 \pm 0.82$ & $<0.48$         & 0.026 &\\	     		     
KPNOTau12     & M9    & 66  & $-0.92 \pm 0.70$ & $<0.42$         & 0.016 &\\	     		     
CFHTBDT1      & M7    & 41  & $-0.29 \pm 0.84$ & $<0.50$         & 0.085 &\\		     	     
CFHTBDT2      & M8    & 47  & $-0.60 \pm 0.80$ & $<0.47$         & 0.036 &\\	     		     
CFHTBDT3      & M9    & 52  & $0.37 \pm 0.77$  & $<0.50$         & 0.016 & \\	     		     
CFHTBDT4      & M7    & 51  & $2.38 \pm 0.75$  & $0.71\pm 0.22$  & 0.085 & $0.4\ldots 2.1$\\	     
J041411+2811  & M6.25 & 74  & $0.91 \pm 0.65$  & $<0.49$         & 0.095 &\\	     		     
J043800+2558  & M7.25 & 59  & $-0.46 \pm 0.80$ & $<0.54$         & 0.065 &\\		     	     
J043814+2611  & M7.25 & 81  & $2.29 \pm 0.75$  & $0.68\pm 0.22$  & 0.065 & $0.4\ldots 2.1$\\	     
J043903+2544  & M7.25 & 59  & $2.86 \pm 0.76$  & $0.85\pm 0.23$  & 0.065 & $0.5\ldots 2.6$\\	     
J044148+2534  & M7.75 & 69  & $2.64 \pm 0.64$  & $0.79\pm 0.19$  & 0.040 & $0.4\ldots 2.4$\\	     
J044427+2512  & M7.25 & 41  & $7.55 \pm 0.89$  & $2.25\pm 0.27$  & 0.065 & $1.2\ldots 6.8$\\	     
J045523+3027  & M6.25 & 29  & $-0.38 \pm 0.93$ & $<0.55$         & 0.095 &\\	     		     
J045749+3015  & M9.25 & 82  & $-0.96 \pm 0.64$ & $<0.48$         & 0.015 &\\	     		     
\enddata
\tablenotetext{a}{Upper limits are based on 2$\sigma$ flux upper limits. Errors correspond to
1$\sigma$ flux uncertainties, and do not include errors in dust temperature, opacity, and distance.}
\tablenotetext{b}{Estimated by varying dust temperature, opacity, and object distance within the
limits discussed in Sect. \ref{dm1}}
\end{deluxetable}

\clearpage

\begin{deluxetable}{lccccccccccc}
\tabletypesize{\scriptsize}
\tablecaption{Model Parameters \label{modelparams}}
\tablewidth{0pt}
\tablehead{
\colhead{Full name} & \colhead{$T_\star$} & \colhead{$R_\star$} & 
\colhead{$M_\star$} & \colhead{$M_d$} & \colhead{$R_d$} &
\colhead{$h_0^{\rm big}$} & \colhead{$h_0^{\rm ISM}$} & \colhead{$f_{\rm ISM}$} & 
\colhead{$\beta$} & \colhead{$i$} & \colhead{$A_V$}\\
\colhead{ } & \colhead{(K)} & \colhead{($R_\odot$)} & \colhead{($M_\odot$)} & 
\colhead{($M_\odot$)} & \colhead{(AU)} &
\colhead{($R_\star$)} & \colhead{($R_\star$)} & \colhead{} & \colhead{ } & \colhead{ } 
& \colhead{ }
}
\tablecolumns{12}
\startdata
CFHTBDT4       & 2900  & 1.2   & 0.10    & $4\times 10^{-4}$   & 100 & 0.02 & 0.04  & 0.03 & 1.15 & 20 & 3.245 \\
               & 2900  & 1.2   & 0.10    & $4\times 10^{-4}$   & 10  & 0.02 & 0.04  & 0.03 & 1.15 & 20 & 3.245 \\
               & 2900  & 1.2   & 0.10    &  1.0                & 1   & 0.02 & 0.035 & $10^{-5}$ & 1.15 & 20 & 3.245 \\
J043814+2611   & 3100  & 1.9   & $\dots$ & $4\times 10^{-4}$   & 100 & 0.03 & 0.06  & 0.03 & 1.25 & 80 & 1.0 \\
J043903+2544   & 2838  & 0.6   & 0.06    & $4\times 10^{-4}$   & 100 & 0.03 & 0.04  & 0.03 & 1.15 & 20 & 0.1 \\
               & 2838  & 0.6   & 0.06    & $4\times 10^{-4}$   & 10  & 0.03 & 0.04  & 0.03 & 1.15 & 20 & 0.1 \\
               & 2838  & 0.6   & 0.06    &  1.0                & 1   & 0.03 & 0.04  & $10^{-5}$ & 1.15 & 20 & 0.1 \\
J044148+2534   & 2838  & 0.42  & 0.04    & $4\times 10^{-4}$   & 100 & 0.05 & 0.07  & 0.03 & 1.15 &20 & 1.5 \\
               & 2838  & 0.42  & 0.04    & $4\times 10^{-4}$   & 10  & 0.04 & 0.06  & 0.03 & 1.15 &20 & 1.5 \\
               & 2838  & 0.42  & 0.04    & 1.0                 & 1   & 0.04 & 0.06  & $10^{-5}$ & 1.15 & 20 &1.5 \\
J044427+2512   & 2900  & 1.4   & 0.05    & $1.2\times 10^{-3}$ & 100 & 0.05 & 0.05  & 0.03 & 1.15 &62.5 & 1.0 \\
               & 2900  & 1.4   & 0.05    & $1.2\times 10^{-3}$ & 10  & 0.05 & 0.05  & 0.03 & 1.15 &62 & 1.0 \\
               & 2900  & 1.4   & 0.05    & 1.                  & 1   & 0.05 & 0.05  & $10^{-5}$ & 1.15 & 59 & 1.0 \\
CFHTBDT2       & 2700  & 0.61  & $\dots$ & $\dots$             & $\dots$ & $\dots$ & $\dots$ & $\dots$ & $\dots$ & $\dots$ & 2.0\\
\enddata
\end{deluxetable}


\clearpage

\clearpage

\clearpage

\clearpage

\end{document}